\documentclass[11pt]{article}
\usepackage[T1]{fontenc}
\usepackage{color}
\usepackage{graphics}

\makeatletter

\providecommand{\LyX}{L\kern-.1667em\lower.25em\hbox{Y}\kern-.125emX\@}

\usepackage{moriond,epsfig}

\def\Journal#1#2#3#4{{#1} {\bf #2}, #3 (#4)}

\def\ApJ{\em Astrophys. J.}
\def\ApJL{\em Astrophys. J. Lett.}
\def\AA{\em Astron. Astrophys.}
\def\AASS{\em Astron. Astrophys. Supp. Ser.}

\def\PRD{{\em Phys. Rev.} D}

\def\PR{\em Phys. Rept.}
\def\NewA{\em New Astron.}
\def\AJ{\em Astron. J.}
\def\CASP{\em Comm. Ap. Sp. Phys.}

\def\be{\begin{equation}}
\def\ee{\end{equation}}
\def\bea{\begin{eqnarray}}
\def\eea{\end{eqnarray}}

\begin{document}

\vspace*{4cm}

\title{SUNYAEV-ZEL'DOVICH EFFECT REVIEW}

\author{F.-X. D\'ESERT}

\address{Laboratoire d'Astrophysique de l'Observatoire de
Grenoble\\ BP53 414 rue de la piscine, Domaine Universitaire de Saint
Martin d'Heres\\ 38041 Grenoble Cedex 9 France}

\maketitle\abstracts{
I review the present observational status of the observations of the SZ effect
due to the Compton scattering of the 3K background on the hot electrons of clusters
of galaxies. I raise the relevant issues from theoretical and X-ray aspects
of the question that challenge the present experiments. Future instruments like
powerful radio interferometers and bolometric cameras should give us access
to the statistics of clusters as well as their internal morphology within the
next ten years. They will provide an approach complementary to space experiments
(XMM, Chandra, Planck) for some fundamental cosmological and large scale structure
issues.}

\section{General Presentation \label{sec:pres}}

The presence\footnote{to be published in the Proceedings of the XXXVth Rencontres de Moriond, Energy Densities in the Universe, Editions
       Frontieres, 2000} of a hot teneous and fully ionised gas (\( T_{e}=10^{8}\, \mathrm{K} \),
\( n_{e}=10^{3}\, \mathrm{m}^{-3} \)) in the intracluster medium was revealed
with the first X-ray measurements toward clusters of galaxies. This gas which
fell in the deep gravitational well of clusters of galaxies and thus heated
up to very high temperature can only cool down via the free-free emission process.
Only in the very center of clusters is the density of the electrons and nuclei
enough for the cooling timescale to be less than the age of the Universe. Another
cooling process exists via inverse Compton scattering on the (cold) cosmic microwave
background.

This secondary cooling is called the Sunyaev-Zel'dovich (hereafter SZ) effect\cite{0}.
This effect preserves the number of CMB photons. If it were a pure scattering
effect without energy change the CMB would not be globally affected. But there
is a net energy gain by the CMB photons in the direction of clusters. The CMB
is thus spectrally distorted. The adimensional Comptonisation parameter \( y \)
measures the SZ distortion:

\begin{equation}
\label{m:eqy}
y={kT_{e}\over m_{e}c^{2}}\sigma _{T}N_{e}\, ,
\end{equation}

where \( T_{e} \), \( m_{e} \), \( N_{e}=\int n_{e}dl \), and \( n_{e} \)
are resp. the electronic temperature, mass, column density and density. The
integral is taken along the line of sight through the cluster. \( \sigma _{T} \)
is the Thomson cross-section. The second (and usually much weaker) SZ effect
called kinetic effect is due to the peculiar cluster velocity \( v_{c} \) and
is measured by:

{\par\centering 
\begin{equation}
\label{m:eqb}
b={v_{r}\over c}\sigma _{T}N_{e}
\end{equation}
\par}

\begin{figure}
{\par\centering \resizebox*{1\textwidth}{!}{\rotatebox{90}{\includegraphics{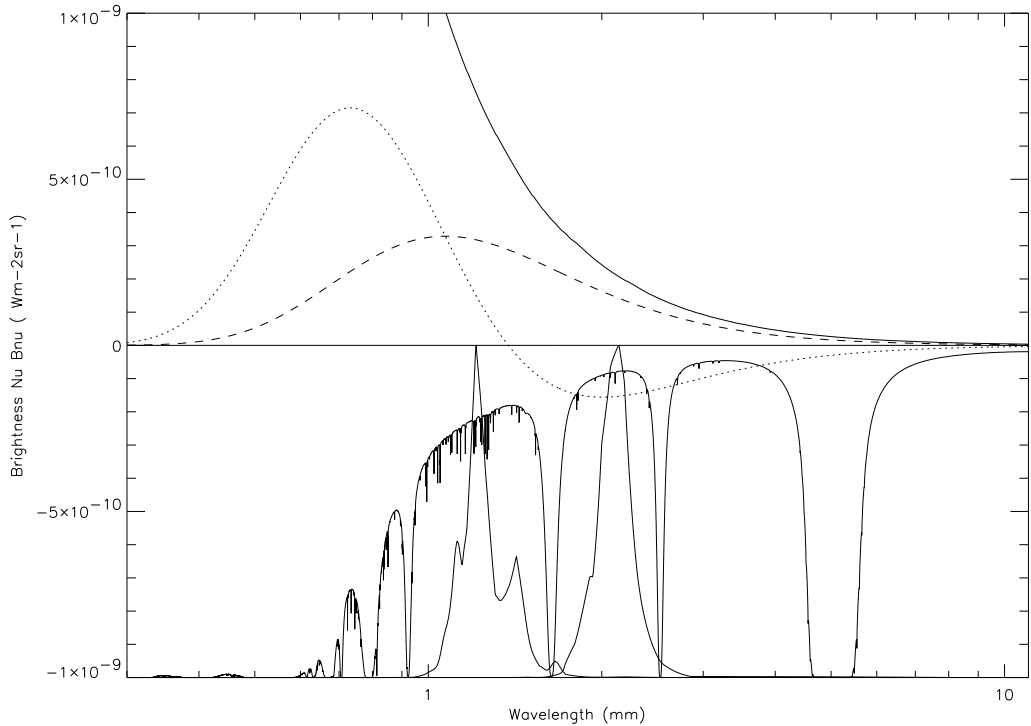}}} \par}

\caption{\label{f:szspectrum}Differential CMB brightness change as a function of wavelength,
towards a cluster with \protect\( y=10^{-4}\protect \) (dotted curve) and \protect\( b=-10^{-4}\protect \)
(dashed curve). The lower panel gives the atmospheric transmission (between
0 and 1) for 3~mm of precipitable water and the two Diabolo instrument bandpasses.
The upper solid curve gives the colour of the atmospheric noise.}
\end{figure}

Figure \ref{f:szspectrum} shows the universal distortion spectrum produced
by the thermal (dots) and kinetic (dashes) SZ effects.

The SZ effect is thus a radio, millimetre and submillimetre phenomenon. The
thermal SZ effect has a very specific spectral signature (always negative for
\( \lambda \leq 1.4\, \mathrm{mm} \)) whereas the kinetic SZ effect is undistinguishable
from the spectrum of the CMB primordial anisotropies. Both SZ effects are brightness
effects which are spectrally independent of redshift in the observer's frame,
contrary to X-ray emission which shows the usual \( (1+z)^{-4} \) brightness
dimming. A well-resolved cluster will show the same SZ effect whether it is
at low or high redshift.

The energy density (which is the focus of this conference) of the CMB is enhanced
towards clusters by the following amount:

{\par\centering 
\begin{equation}
\label{m:eqc}
E=4y\sigma _{S}T^{4}_{CMB}
\end{equation}
\par}

The opacity \( \tau =\sigma _{T}N_{e} \) and comptonisation parameter \( y \)
are of the order of a few \( 10^{-2} \) and \( 10^{-4} \) resp. in the richest
clusters which therefore make the SZ effect a relatively small and linear distortion. 

It has recently been acknowledged that one cannot neglect relativistic corrections
to the (non-relativistic) universal spectral template, shown in Fig.~\ref{f:szspectrum},
and independent on redshift. A complete review of the SZ effect is clearly beyond
this presentation. An exhaustive recent review was made by Birkinshaw\cite{1}.

\section{Observational status\label{Se:Obs}}

\subsection{Astrophysical and Cosmological objectives\label{Sse:As}}

\subsubsection{Witness the 3K remoteness}

With several clusters firmly detected at a redshift up to about 0.5, the 3~K
background cannot have an origin in the local Universe. There are not so many
direct probes of the presence of the CMB at high redshift.

\subsubsection{Measure the hot cluster gas distribution }

The SZ effect directly measures the hot electron pressure along the line of
sight. If one assumes a constant gas temperature, an electronic density following
a King profile (with \( r \) the radius from the cluster center, and \( a \)
the core radius) is often assumed:

\begin{equation}
\label{eq:ne}
n_{e}=n_{e0}(1+(r/a)^{2})^{-3\beta /2},
\end{equation}

and will produce a SZ angular distribution

\begin{equation}
\label{eq:ytheta}
y_{SZ}=y_{0}(1+(\theta /\theta _{a})^{2})^{-3\beta /2+1/2}.
\end{equation}

The measurements are therefore linearly linked to the density of baryonic matter.

\subsubsection{Total mass of gas \textcolor{black}{\small }\small }

\textcolor{black}{\small The quantity that is directly measured in a given experiment
is really the brightness integrated over the instrument beam, something that
can be loosely defined as \( Y=\int _{beam}yd\Omega  \). The big virtue of
SZ measurements is to give an easy access which is weakly dependent on redshift
to the total gas mass of the cluster in the beam:} 
\begin{equation}
\label{eq:mgas}
M_{g}\simeq 2\times 10^{13}M_{_{\odot }}({T_{g}\over 10\, \mathrm{keV}})^{-1}{Y\over 10^{-5}\mathrm{arcmin}^{2}}f(z),
\end{equation}
 where the redshift function \( f \) depends on the cosmological parameters
(through the angular-diameter redshift dependence). The following table (computed
for a critical standard model without cosmological constant) shows a subtle
dependence of the measured SZ effect with redshift.

\vspace{0.3cm}
{\centering \begin{tabular}{|c|c|}
\hline 
\( z \)&
\( f\times10 ^{3} \)\\
\hline 
\hline 
0.1&
2\\
\hline 
0.3&
9\\
\hline 
1.0&
20\\
\hline 
\end{tabular}\par}
\vspace{0.3cm}

\subsubsection{Total gravitational mass }

Through the hydrostatic equation and from the gas pressure profile, one can
deduce the total cluster mass \( M_{T} \). Although this has been done with
X-ray measurements, the SZ effect could in principle be used for a rather direct
measurement of the total mass. It would be valuable, when precise measurements
become available, to reassess the baryonic crisis: \( {M_{g}/M_{T}}\neq \Omega _{b}/\Omega _{0} \).

\subsubsection{Peculiar radial velocity }

The kinetic SZ effect is a 10 times weaker effect that the thermal effect. Accurate
measurements of it in many cluster could in principle probe the large scale
velocity field in the distant Universe. The CMB primary anistropies are in that
case a `pollution' to these measurements which could be attempted by the Planck
mission (Aghanim\cite{10}). Ground-based attempts have so far provided upper
limits (Holzapfel et al.\cite{7})

\subsubsection{\protect\( H_{0}\protect \) , \protect\( q_{0}\protect \) measurement}

The measurements of \( \int T_{e}n_{e}dl \) with the SZ effect and \( \int n_{e}^{2}dl \)
and \( T_{e} \) with X-ray space observations yield an estimate of the true
physical depth of the cluster. Assuming the cluster is spherical, this quantity
can be compared with the angular size of the cluster and its redshift to give
\( H_{0} \). The weak dependence of that result on \( q_{0} \) has been analysed,
but the prospect of a serious measurement of it is marred by cluster evolution
(see below). Measurements of the Hubble constant is clearly within reach, once
systematic effects are well understood over a statistically significant sample
of observed (local) clusters. The complementarity of the SZ effect with XMM-Newton
and Chandra is obvious in that respect. This is one of the most important cosmological
targets for SZ measurements. The second one is:

\subsubsection{SZ Cluster number counts }

Having SZ surveys over large area could provide a rather unbiased measurement
of the cluster number counts. Optical and X-ray surveys have been known to be
biased by chance alignments and resolution \& surface brightness limits respectively.
The weak dependence on redshift of the SZ effect (Eq.~\ref{eq:mgas}) makes
a (costly) survey quite attractive for two reasons: 

\begin{enumerate}
\item to determine the exact number density of local clusters (say of redshift less
than 0.3) and the mass distribution function
\item to measure the evolution of this density and search for high redshift clusters.
\end{enumerate}
\begin{figure}
{\par\centering \includegraphics{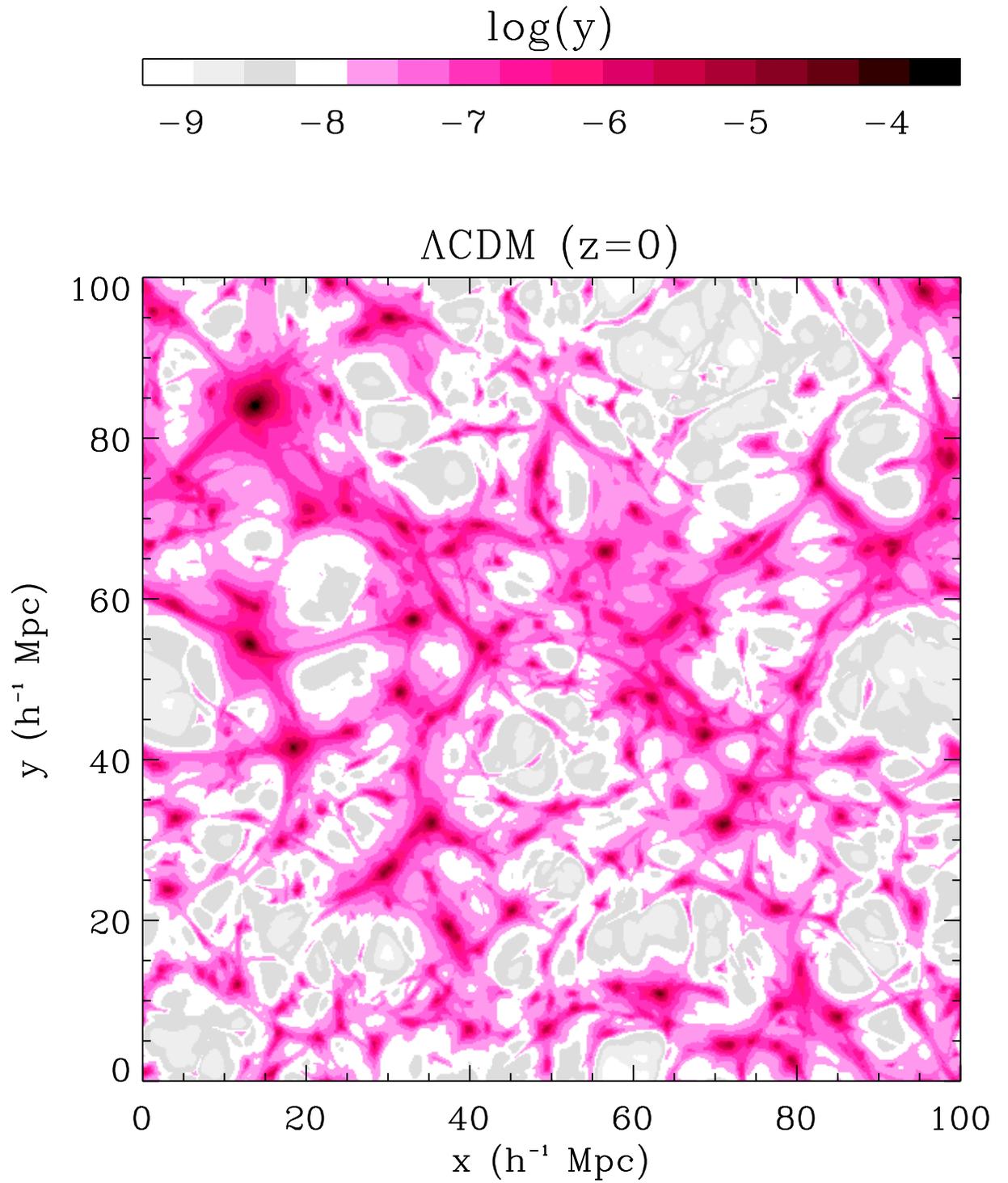} \par}

\caption{A SZ map produced by a simulated cube of Universe (Refregier
et al.$^{10}$).  Only the darkest spots corresponding to cluster
cores can be measured now. Notice the SZ spider web like network that
crosses the sky.}
\end{figure}

\subsubsection{Structured Matter Energy injection into the CMB }

The average comptonisation parameter on any line of sight is about \( y_{}\sim 10^{-6} \).
From Eq.~\ref{m:eqc}, the energy injection into the Universe by large scale
non-linear structure formation is of the order of at most \( 10^{-5} \) of
the energy in the CMB itself.

\subsubsection{Search for distant clusters : \protect\( z>1\protect \)}

The search of SZ effect without the a-priori of X-ray maps have so far led to
the as yet unconfirmed detection of two radio extended brightness decrements
(Richards et al.\cite{11}, Saunders et al.\cite{12}). The secure detection
of just few clusters at redshift above 1 would severely endanger models of the
Universe with a critical matter density parameter (Bartlett et al.\cite{13}).

\subsection{The three observational techniques\label{Sse:Tech}}

\subsubsection{Single dish radio telescopes}

Pionneered by Birkinshaw and collaborators, this was the first standard technique
to successfully measure the SZ effect. It was and remains the best adapted technique
for local clusters with a very large extent (e.g. Coma: Herbig et al.\cite{15})

\subsubsection{Radio interferometers}

The most sensitive detections have been provided by radio interferometers which
are less affected by sidelobe effects and can remove point sources by using
the long baseline measurements. Moreover they benefit from large integration
times (several months per year). The Ryle telescope (Saunders and Jones, this
conference) in England and the OVRO-BIMA interferometer fitted (perverted) with
radio receivers (Carlstrom et al.\cite{9}) have obtained very sensitive maps
of now tens of clusters with arcminute resolution at 15 and 30~GHz frequencies.

\subsubsection{Bolometer photometers}

The SuZie experiment provided the first detection of the SZ effect in the millimeter
domain at the CSO 10m telescope (Wilbanks et al.\cite{8}, Holzapfel et al.\cite{7}).
It is made of 6 bolometers at a wavelength of 2~mm. By using a drift scan technique
whereby fixing the telescope in local coordinates, the cluster drifts through
the detectors with the Earth's diurnal motion, hence avoiding sidelobe effects
and microphonic noise. The cluster A~2163 (the second X-ray brightest known
cluster) was later detected at 1mm using the SuZie experiment with different
filters and in the submillimeter domain by the balloon borne photometer PRONAOS-SPM
(Lamarre et al.\cite{6}) showing for the first time the change of sign of the
SZ effect (see Fig.~\ref{f:szspectrum}). Although in the (sub)millimeter domain,
the point sources should not contaminate the SZ measurements so much, interstellar
dust thermal emission must be dealt with to correct the measurements. Another
limitation comes from sky noise: the water vapour is inhomogeneous in the atmosphere.
Its emission in the telescope is variable in time and angle and frequency (see
Fig.~\ref{f:szspectrum}). The Diabolo experiment (Benoit et al.\cite{3}) uses
a dichroic beam splitter with six 0.1~K bolometers in order to simultaneously
measure and hence subtract the water vapour emission at 1.2~mm (where the SZ
effect is almost vanishing) from the SZ measurement at 2.1~mm. At the IRAM 30~m
telescope, it provided the highest angular resolution of the SZ effect on several
clusters with 30 arcsecond beam in 1995 (Désert et al.\cite{5}) and 22 arcsecond
beam since 1997 (Pointecouteau et al.\cite{4} and this conference).

\section{The next SZ ten years\label{Se:Ten}}

The sensitivity of present ground-based detectors is quite close to photon noise
limits, typically an effective value \( y(1\sigma )\sim 10^{-4}\mathrm{hour}^{-\frac{1}{2}} \)
for a single bolometer. This is also typical for interferometers. All 3 kinds
of observing techniques are also currently limited by the range of angular scales
that can be measured, whereas the angular distribution of clusters of galaxies
is widespread, from core radii to Abell radii, along with substructure scales.
We can see that the main goals of SZ observations in the next years are:

\begin{enumerate}
\item Improve the statistics on \( H_{0} \) measurements
\item Number counts of SZ clusters and high redshift cluster search
\item Detailed analysis of individual clusters
\end{enumerate}
For point 1 and 2, interferometers are clearly very promising. Dedicated radio
telescope arrays which are currently being built with modest size telescopes
aim at covering a large range of angular scales (in particular the shortest
baselines). By covering a large frequency bandwidth and by using
smaller telescopes (!), improvements in detectivity
could be better by as much as a factor of 1000 in the next 3 years (e.g. the
AMI project: Jones, this conference). For point 3, bolometer arrays (with hundred
to thousands of pixels) should bring a clear multiplex advantage over existing
technology\cite{16}. They will give unprecedented high angular resolution maps
in the near future (20~arcseconds), that will be useful to study the detailed
angular distribution of clusters (whether at low or high redshift) to unravel
the cosmogony of large scale structures. In that respect, the structure of high
redsihft clusters (e.g. Ebeling et al. \cite{18}) which is far from smooth
is interesting for cluster formation scenarios but may prove a show stopper
for \( q_{0} \) and other second order effect measurements.

Future space missions will provide a different perspective. Planck will give
an unbiased catalog of at least several thousands of SZ sources. MAP, although
not sensitive to individual clusters, may still see some signal by correlation
with large scale structures as seen in the optical (Refregier et al.\cite{17}).

The interpretation of SZ data is depending on the quality of other data, and
vice-versa. The arrival of Chandra and XMM-Newton is a strong incentive for
improving SZ measurements. Comparison with visible and near-infrared data obtained
by large telescopes (substructures and weak lensing) is also very valuable.

We have clearly moved from detection experiments towards a powerful tool for
the study of clusters. A global approach, using SZ observations but also other
wavelengths, is a must for the understanding of clusters of galaxies.

\section*{Acknowledgments}

We wish to thank the organisers for such a pleasant and lively forum of discussions
of which the SZ effect was one of the foci. We thank A. Refregier for allowing
us to show the large scale SZ simulated map.

\end{document}